\documentclass[12pt]{article}
\usepackage[utf8]{inputenc}
\usepackage[english]{babel}
\usepackage{epsfig}
\usepackage{amsmath}
\usepackage{amssymb}
\usepackage{color}
\usepackage{bbm}
\usepackage{hyperref}
\usepackage{physics}
\usepackage{graphicx}
\usepackage{xcolor,colortbl}
\usepackage{setspace}
\usepackage[margin=50pt]{geometry}
\usepackage{slashed}

\usepackage{cite}

\newcommand{\be}{\begin{equation}}
\newcommand{\ee}{\end{equation}}
\newcommand{\ba}{\begin{eqnarray}}
\newcommand{\ea}{\end{eqnarray}}
\newcommand{\bs}{\begin{subequations}}
\newcommand{\es}{\end{subequations}}

\usepackage{subcaption}

\allowdisplaybreaks

\begin{document}
\large

\title{Complete set of bounded-from-below conditions for the $A_4$-symmetric
three-Higgs-doublet model}

\author{
  Lu\'is Lavoura\thanks{E-mail:
    \tt balio@cftp.tecnico.ulisboa.pt}
    \ and\ 
  Andr\'e Milagre\thanks{E-mail:
    \tt andre.milagre@tecnico.ulisboa.pt}
  \
  \\*[3mm]
  \small Universidade de Lisboa, Instituto Superior T\'ecnico, CFTP, \\
  \small Av.~Rovisco~Pais~1, 1049-001~Lisboa, Portugal
}

\maketitle

\begin{abstract}
  We derive bounded-from-below \emph{necessary} conditions
  for the scalar potential of an $A_4$-invariant three-Higgs-doublet model.
  The field {\it Ans\"atze}\/ that we have used
  were devised with the help of Claude.
  Testing our
  set of conditions on $10^6$ random potentials,
  that in parallel we minimize by brute force using the 
  {\tt Mathematica} package {\tt StableWein},
  we find that the set seems to be \emph{necessary and sufficient}
  for boundedness-from-below.
\end{abstract}

\section{Introduction}

We consider the $SU(2) \times U(1)$ gauge model
with scalar sector consisting of three $SU(2)$ doublets with equal hypercharge;
this is called the three-Higgs-doublet model (3HDM).
Let $\chi_k$ ($k=1,\,2,\,3$) denote the doublets.
We study the version of the 3HDM where,
besides the $SU(2)\times U(1)$ gauge symmetry,
there is also invariance under the discrete non-Abelian group $A_4$
generated by the transformations
\bs
\begin{eqnarray}
  \mathcal{T}_1:&& \chi_2 \to -\chi_2,\ \chi_3 \to -\chi_3;
  \\
  \mathcal{T}_2:&& \chi_1 \to \chi_2 \to \chi_3 \to \chi_1.
\end{eqnarray}
\es
Such a model has been used 
to address the flavour puzzle~\cite{
  Lavoura:2007dw,
  Machado:2010uc,
  deAdelhartToorop:2010jxh,
  deAdelhartToorop:2010nki,
  Ma:2010gs,
  GonzalezFelipe:2013xok,
  GonzalezFelipe:2014mcf,
  Bree:2023ojl},
to generate neutrino masses~\cite{
  Ma:2001dn,
  Ma:2004zv,
  Morisi:2009sc,
  GonzalezFelipe:2013yhh,
  Ma:2015fpa,
  Izawa:2022viu,
  Dziewit:2025xun,
  Chulia:2025nsa},
as an explanation for dark matter~\cite{
  Hirsch:2010ru,
  Esteves:2010sh,
  Meloni:2010sk,
  Boucenna:2011tj,
  deAdelhartToorop:2011ad,
  deMedeirosVarzielas:2022kbj,
  Bonilla:2023pna,
  Kumar:2024zfb},
amongst other applications~\cite{
  Heeck:2014qea,
  Bhattacharyya:2014oka,
  deMedeirosVarzielas:2015ybd,
  Branco:2015bfb,
  deMedeirosVarzielas:2016rii,
  Pramanick:2017wry,
  Chakrabarty:2018yoy}.

Using
\begin{equation}
	A_{kl}:= \chi_k^\dagger \chi_l,
\end{equation}
we write the most general renormalizable 
$SU(2) \times U(1) \times A_4$-symmetric scalar potential (SP)
as $V = V_2 + V_4$,
where $V_2 = \mu \left( A_{11} + A_{22} + A_{33} \right)$ 
and~\cite{Degee:2012sk}
\bs
\label{SP_1}
\ba
	V_4 &=& 
	\frac{\Lambda_0+\Lambda_3}{3}
		\left( A_{11}^2 + A_{22}^2 + A_{33}^2\right)
	+\frac{2\Lambda_0-\Lambda_3}{3}
		\left( A_{11}A_{22} + A_{22}A_{33} + A_{33}A_{11}\right)\\
	&& +\  \frac{\Lambda_1+\Lambda_2}{2} 
		\left(|A_{12}|^2 + |A_{23}|^2 + |A_{31}|^2\right)
	+ \frac{\Lambda_1-\Lambda_2}{2}\,
	\Re\left( A_{12}^2 + A_{23}^2 + A_{31}^2 \right) \hspace*{7mm} \\
	&&+\ \frac{\Lambda_4}{2}\,
        \Im \left( A_{12}^2 + A_{23}^2 + A_{31}^2 \right).
\ea\es
Both $\mu$ and $\Lambda_{0,1,2,3,4}$ are real.
In the special case where $\Lambda_4 = 0$,
the potential is $S_4$-symmetric~\cite{Degee:2012sk,Ivanov:2020jra}.

In this work,
we are interested in the conditions that
guarantee that an SP with $V_4$ given by Eq.~\eqref{SP_1}
is bounded from below (BFB).\footnote{If the SP is not BFB,
then the theory has no vacuum state, and it is of no use.}
That is,
we look for conditions that ensure that
\emph{$V_4 \ge 0$ for all possible field configurations}.\footnote{For
boundedness from below (BnessFB) only $V_4$ matters.
$V_2$ may include a soft breaking of the symmetry $A_4$.}\footnote{Some authors
prefer to define BnessFB as $V_4 > 0$ instead of $V_4 \ge 0$.
If one wants to use that definition,
then one may still use the results in this paper,
with all $\ge$ and $\le$ signs substituted by $>$ and $<$,
respectively.}
To that end, it is convenient to rewrite $V_4$ as
\be
\label{v4}
V_4 = 
N^2 \left(
\frac{\Lambda_0 + \Lambda_3}{3} - \Lambda_3\, x
+ \frac{\Lambda_1 + \Lambda_2}{2}\, x'
+ \frac{\Lambda_1 - \Lambda_2}{2}\, y
+ \frac{\Lambda_4}{2}\, z 
\right),
\ee
where
\bs
\label{Nxyz}
\begin{eqnarray}
	N &\equiv& A_{11} + A_{22} + A_{33},\\[2pt]
	x &\equiv& \frac{A_{11}A_{22} + A_{22}A_{33} + A_{33}A_{11}}{N^2},\\[4pt]
	x' &\equiv& \frac{|A_{12}|^2 + |A_{23}|^2 + |A_{31}|^2}{N^2},
        \\[4pt]
	y + i z &\equiv& \frac{A_{12}^2 + A_{23}^2 + A_{31}^2}{N^2}.
\end{eqnarray}
\es
The set of values that $\left( x, x', y, z \right)$ may take
is called the orbit space.

Notice that $\Lambda_0$ appears only as an additive constant in Eq.~\eqref{v4},
\textit{i.e.}\ the $\Lambda_0$ term is independent from the orbit-space point.
Therefore,
requiring $V_4 \ge 0$ at every point of orbit space
implies BFB conditions of the form~\cite{Carrolo:2022oyg}
\begin{equation}
\label{gen_cond}
	\Lambda_0 + 
	\Sigma\left( \Lambda_1, \Lambda_2, \Lambda_3, \Lambda_4\right) \ge 0.
\end{equation}
In sections~\ref{s:conf} and~\ref{s:lines} we derive various 
$\Sigma\left( \Lambda_1, \Lambda_2, \Lambda_3, \Lambda_4\right)$
that provide a set of \emph{necessary} BFB conditions.
In section~\ref{s:conc} we suggest that our set of necessary conditions
actually is \emph{necessary and sufficient} (n\&s) for BnessFB.
Section~\ref{s:s4} is devoted to the $S_4$-symmetric case.

\section{Vertices}
\label{s:conf}

We say that a configuration of the scalar doublets is `neutral'
if one may use an $SU(2)$ transformation
to nullify all the upper components of the doublets.
If a configuration of the doublets is not neutral
then we call it `charge-breaking'.
In neutral configurations the doublets behave like singlets,
meaning that $\left| A_{ij} \right|^2 = A_{ii} A_{jj}\ \forall i, j$.
Thus, in the notation of~\eqref{v4},
the formal definition of a neutral configuration is $x=x'$.

\subsection{Neutral vertices}
\label{ss:neutral}

\begin{figure}[h!]
\centering
  \includegraphics[width=0.4\textwidth]{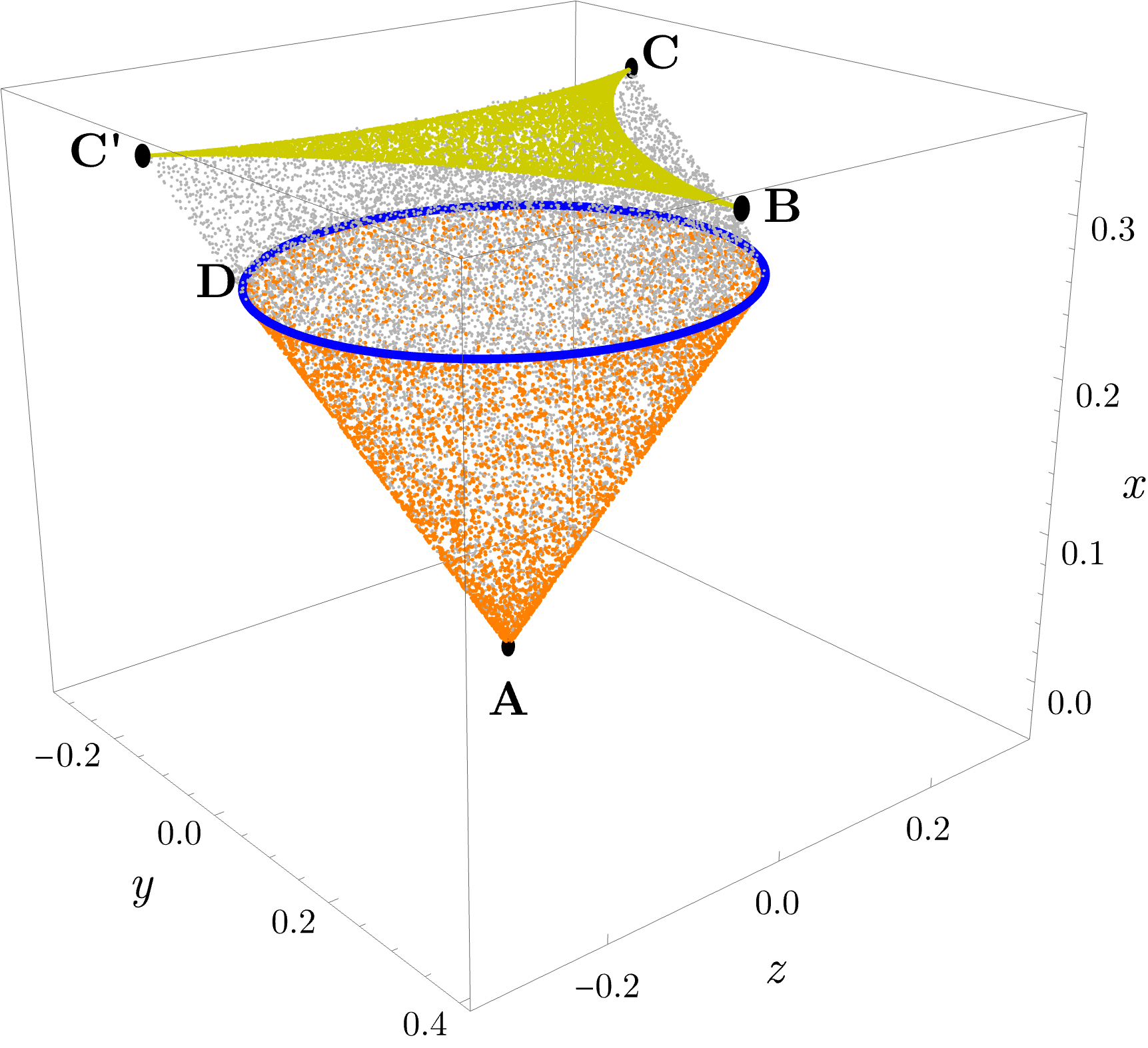}
  \caption{The points of orbit space with $x=x'$.}
  \label{fig_neutral}
\end{figure}

The possible neutral vacua of the $A_4$-symmetric 3HDM
have been found in Refs.~\cite{Degee:2012sk,Carrolo:2022oyg}.
They are
\begin{itemize}
\item point $A$:
  $\left( x, x', y, z \right) = \left(0, 0, 0, 0\right)$;
\item point $B$:
  $\left( x, x', y, z \right) = \frac{1}{3} \left(1, 1, 1, 0\right)$;
\item point $C$:
  $\left( x, x', y, z \right) = \frac{1}{6} \left( 2, 2, -1, \sqrt3 \right)$;
\item point $C'$:
  $\left( x, x', y, z \right) = \frac{1}{6} \left( 2, 2, -1, -\sqrt3 \right)$;
\item points in circumference $D$:
  $\left( x, x', y, z \right) = \frac{1}{4}
  \left( 1, 1, \cos{2 \alpha}, \sin{2 \alpha} \right)$,
  with $\alpha\in[0,\pi]$.
\end{itemize}
Figure~\ref{fig_neutral} depicts the orbit space sliced along
the hyperplane $x=x'$.
The points $A$,
$B$,
$C$,
and~$C'$ are marked, and
the circumference $D$ is drawn in blue.

The points and circumference may be obtained,
for example,
by using the configurations of doublets\footnote{
We do not care about the overall normalization of the doublets because $x$,
$x'$,
$y$,
and $z$ are invariant under a simultaneous rescaling of all three $\chi_k$.}
\bs
\ba
A: & &
\chi_1 = \left( \begin{array}{c} 0 \\ \varrho^2 \end{array} \right),
\quad
\chi_2 = \chi_3 = \left( \begin{array}{c} 0 \\ 0 \end{array} \right);
\label{A} \\
B: & & \chi_1 = \chi_2 = \chi_3 =
\left( \begin{array}{c} 0 \\ \varrho \end{array} \right);
\label{B} \\
C: & & \chi_1
= \left( \begin{array}{c} \omega^2 \varrho \\ 0 \end{array} \right),
\quad
\chi_2 = \left( \begin{array}{c} \omega \varrho \\ 0 \end{array} \right),
\quad
\chi_3 = \left( \begin{array}{c} \varrho \\ 0 \end{array} \right);
\label{C} \\
C': & &
\chi_1 = \left( \begin{array}{c} \omega \varrho \\ 0 \end{array} \right),
\quad
\chi_2 =
\left( \begin{array}{c} \omega^2 \varrho \\ 0 \end{array} \right),
\quad
\chi_3 = \left( \begin{array}{c} \varrho \\ 0 \end{array} \right);
\label{C'} \\
D: & & \chi_1 = \left( \begin{array}{c} 0 \\ 0 \end{array} \right), \quad
\chi_2 = \left( \begin{array}{c} \varrho \\ 0 \end{array} \right), \quad
\chi_3 = \left( \begin{array}{c} e^{i\alpha} \varrho \\ 0 \end{array} \right),
\label{D}
\ea
\es
where $\varrho = \sqrt{2}$ and $\omega = e^{2i\pi/3}$.

As stated, for BnessFB, $V_4$ cannot be negative at any of these points.
The resulting necessary BFB conditions happen to be n\&s
if one restricts oneself to
neutral configurations of doublets~\cite{Ivanov:2020jra, Buskin:2021eig}.
Written in the form of inequality~\eqref{gen_cond},
they have $\Sigma\left( \Lambda_1, \Lambda_2, \Lambda_3, \Lambda_4\right)$
equal to
\bs
\begin{eqnarray}
  \Lambda_3 &\equiv& \Sigma_A,\\
  \Lambda_1 &\equiv& \Sigma_B,\\
  \frac{\Lambda_1 + 3\Lambda_2 + \sqrt3 \Lambda_4}{4} &\equiv& \Sigma_C ,\\
  \frac{\Lambda_1 + 3\Lambda_2 - \sqrt3 \Lambda_4}{4} &\equiv& \Sigma_{C'},\\
  \frac{\Lambda_3}{4} + \frac{3}{8} \left[ \Lambda_1 + \Lambda_2
      + (\Lambda_1 - \Lambda_2) \cos 2\alpha
      + \Lambda_4 \sin 2\alpha\right] &\equiv&
  \Sigma_{D} \left( \alpha \right),
\end{eqnarray}
\es
respectively.
Since $\Lambda_0 + \Sigma_{D} \left( \alpha \right) \ge 0\ \forall \alpha$,
one may minimize $\Sigma_{D} \left( \alpha \right)$ with respect to $\alpha$
and obtain~\cite{Ivanov:2020jra}
\be
\frac{\Lambda_3}{4} + \frac{3}{8} \left( \Lambda_1 + \Lambda_2 - R \right)
\equiv \Sigma_D,
\ee
where
\be
R := \sqrt{(\Lambda_1 - \Lambda_2)^2 + \Lambda_4^2}.
\ee

\subsection{Charge-breaking vertices}
\label{ss:charged}

\begin{figure}[h!]
  \centering
  \includegraphics[width=0.5\textwidth]{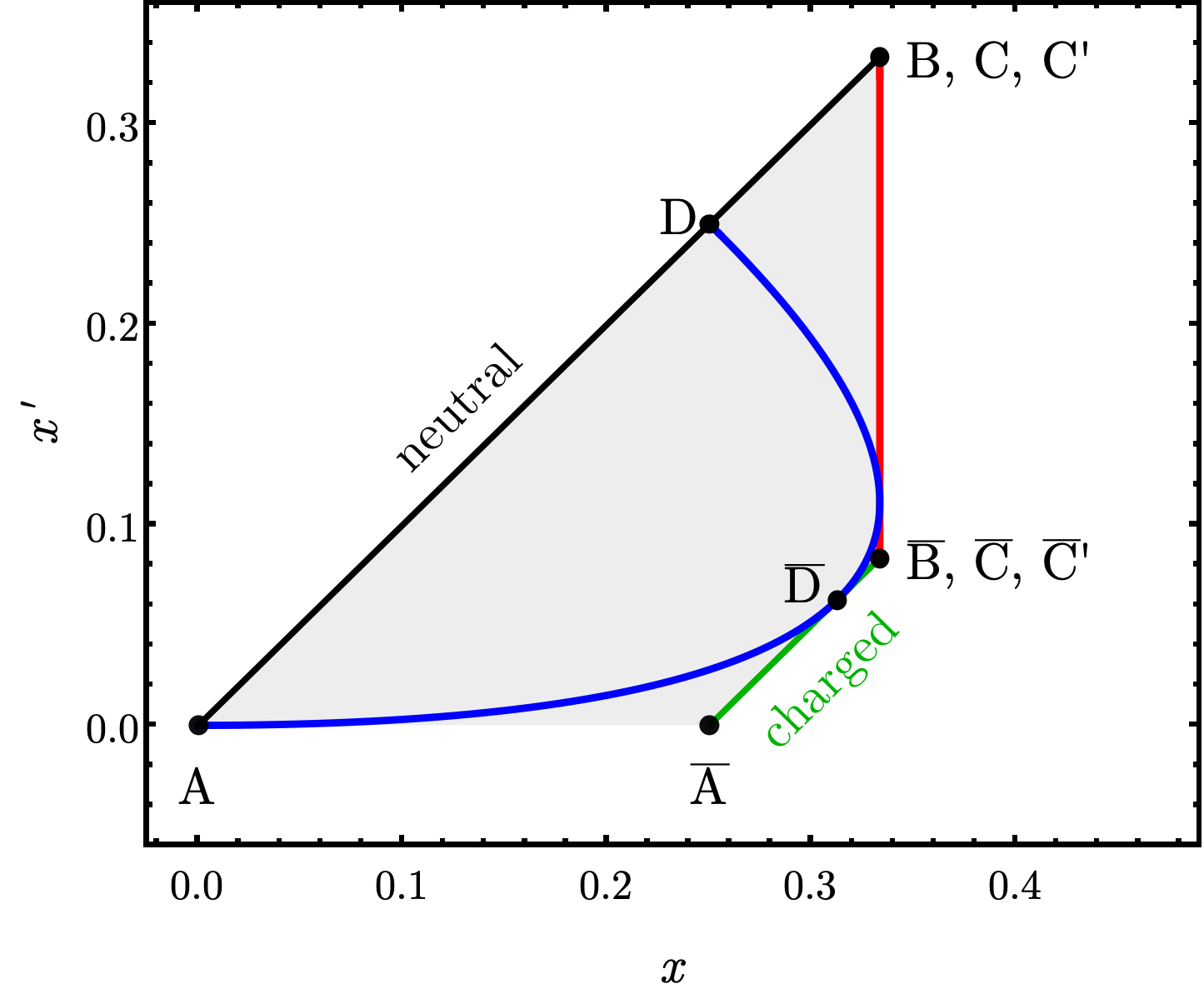}
  \caption{Orbit space projected on the $(x, x')$ plane.
    The black line contains neutral configurations.
    The green line contains the charge-breaking vertices.
    Red line: the three $t$-paths (they all coincide in this projection).
  Blue line: the $q$-paths (they all coincide in this projection).}
  \label{f:paths}
\end{figure}

In Fig.~\ref{f:paths} we show the orbit space projected
onto the $(x, x')$ plane.
The neutral configurations lie on the black line where $x' = x$. 
One sees that there are additional vertices on the green line $x' = x - 1/4$. 
Their coordinates are
\begin{itemize}
\item point $\overline{A}$: $\left( x, x', y, z \right)
  = \frac{1}{4} \left( 1, 0, 0, 0\right)$;
\item point $\overline{B}$: $\left( x, x', y, z \right)
  = \frac{1}{12} \left( 4, 1, 1, 0\right)$;
\item point $\overline{C}$: $\left( x, x', y, z \right)
  = \frac{1}{24} \left( 8, 2, -1, \sqrt3 \right)$
\item point $\overline{C'}$: $\left( x, x', y, z \right)
  = \frac{1}{24} \left( 8, 2, -1, -\sqrt3 \right)$
\item circumference $\overline{D}$: $\left( x, x', y, z \right) =
  \frac{1}{16} \left( 5, 1, \cos{2 \alpha}, \sin{2 \alpha} \right)$.
\end{itemize}
As first noticed for the $S_4$-symmetric model in Ref.~\cite{Ivanov:2020jra}, 
these vertices can only be reached through charge-breaking field 
configurations. Examples of such configurations are
\bs
\ba
\overline{A}: & &
\chi_1 = \left( \begin{array}{c} 0 \\ 1 \end{array} \right), \quad
\chi_2 = \left( \begin{array}{c} 1 \\ 0 \end{array} \right), \quad
\chi_3 = \left( \begin{array}{c} 0 \\ 0 \end{array} \right);
\label{oA} \\
\overline{B}: & &
\chi_1 = \left( \begin{array}{c} \omega^2 \\ \omega \end{array} \right), \quad
\chi_2 = \left( \begin{array}{c} \omega \\ \omega^2 \end{array} \right), \quad
\chi_3  = \left( \begin{array}{c} 1 \\ 1 \end{array} \right);
\label{oB} \\
\overline{C}: & &
\chi_1 = \left( \begin{array}{c} \omega \\ 1 \end{array} \right), \quad
\chi_2 = \left( \begin{array}{c} \omega^2 \\ 1 \end{array} \right), \quad
\chi_3 = \left( \begin{array}{c} 1 \\ 1 \end{array} \right);
\label{oC} \\
\overline{C'}: & &
\chi_1 = \left( \begin{array}{c} \omega^2 \\ 1 \end{array} \right), \quad
\chi_2 = \left( \begin{array}{c} \omega \\ 1 \end{array} \right), \quad
\chi_3 = \left( \begin{array}{c} 1 \\ 1 \end{array} \right);
\label{oC'} \\
\overline{D}: & &
\chi_1 = \left( \begin{array}{c} 0 \\ \varrho \end{array} \right), \quad
\chi_2 = \left( \begin{array}{c} 1 \\ 0 \end{array} \right), \quad
\chi_3 = \left( \begin{array}{c} e^{i\alpha} \\ 0 \end{array} \right).
\label{oD}
\ea
\es

As the naming suggests,
each charge-breaking configuration has an associated neutral configuration:
at the charge-breaking configuration,
$\Sigma \left( \Lambda_1, \Lambda_2, \Lambda_3, \Lambda_4 \right)$
is one quarter of its value at the corresponding
neutral configuration.
Thus,
\begin{equation}
  \Sigma_{\overline{A}} = \frac{\Sigma_A}{4},
  \qquad
  \Sigma_{\overline{B}} = \frac{\Sigma_B}{4},
  \qquad
  \Sigma_{\overline{C}} = \frac{\Sigma_C}{4},
  \qquad
  \Sigma_{\overline{C'}} = \frac{\Sigma_{C'}}{4},
  \qquad
  \Sigma_{\overline{D}} = \frac{\Sigma_D}{4}.
\end{equation}

\section{Connecting the neutral and charge-breaking vertices}
\label{s:lines}

\subsection{Mathematical preliminary}

Consider the function $f:\:\mathbb{R}\to\mathbb{R}$ defined as
\be
f(u) = a u^2 + 2 k u + c,
\ee
where the coefficients $a$, $k$,
and $c$ are real.
Suppose one wants to guarantee that $f(u) \ge 0$ in the whole domain
$u \in \left[ -1, 1 \right]$.
This, of course, implies that both 
$f(-1) \ge 0$ and $f(1) \ge 0$.
Thus, one must enforce
\be
a - 2 k + c \ge 0, \qquad a + 2 k + c \ge 0.
\ee
Additionally,
if $f(u)$ has a \emph{minimum} at the point $u=u_0$,
and if $u_0$ happens to be inside the interval $\left] -1, 1 \right[$,
then one must require $f(u_0) \ge 0$.
This means that
\be
\mathrm{if}\ a > 0,\ a - k > 0,\ \mathrm{and}\ a + k > 0,\ 
\mathrm{then}\ c - \frac{k^2}{a} \ge 0.
\ee

\subsection{The $t$-paths}

\subsubsection{$t$-path~1}

One may connect vertices $B$,
$\overline{C}$,
and $C'$ at Eqs.~\eqref{B},
\eqref{oC},
and~\eqref{C'},
respectively,
through the \textit{Ansatz}
\be
\label{nfo}
\chi_1 = \left( \begin{array}{c} \omega \sqrt{1 + u} \\*[2mm] \sqrt{1 - u}
\end{array} \right),
\quad
\chi_2 = \left( \begin{array}{c} \omega^2 \sqrt{1 + u} \\*[2mm] \sqrt{1 - u}
\end{array} \right),
\quad
\chi_3 = \left( \begin{array}{c} \sqrt{1 + u} \\*[2mm] \sqrt{1 - u}
\end{array} \right),
\ee
where $u \in \left[ -1, 1 \right]$.
Indeed,
when $u = -1$,
Eq.~\eqref{nfo} reproduces Eq.~\eqref{B};
when $u = 0$,
it reproduces Eq.~\eqref{oC};
and when $u = 1$,
it reproduces Eq.~\eqref{C'}.
Now,
configuration~\eqref{nfo} produces $A_{11} = A_{22} = A_{33} = 2$
and $A_{12} = A_{23} = A_{31} = - \omega^2 \left( 1 + i u \sqrt{3} \right)$,
hence
\bs
\label{jfp}
\ba
N &=& 6, \\
x &=& \frac{1}{3}, \label{x1} \\
x' &=& \frac{3 u^2 + 1}{12}, \label{x'1} \\
y &=& \frac{3 u^2 - 6 u - 1}{24}, \label{y1} \\
z &=& \frac{\sqrt{3} \left( - 3 u^2 - 2 u + 1 \right)}{24}. \label{z1}
\ea
\es
We call the orbit-space line given
(in parametric form)
by Eqs.~\eqref{x1}--\eqref{z1} the $t$-path~1.

Plugging the values~\eqref{jfp} into Eq.~\eqref{v4}
one obtains $V_4 = 12 \left( a u^2 + 2 k u + c \right)$,
with
\bs
\ba
a &=& \frac{9 \Lambda_1 + 3 \Lambda_2 - 3 \sqrt{3}\, \Lambda_4}{16},
\\
k &=& \frac{- 3 \Lambda_1 + 3 \Lambda_2 - \sqrt{3}\, \Lambda_4}{16},
\\
c &=& \Lambda_0 + \frac{\Lambda_1 + 3 \Lambda_2 + \sqrt{3}\, \Lambda_4}{16}.
\ea
\es
So,
according to the previous subsection,
one must demand
\be
\Lambda_0 + \Sigma_B \ge 0, \qquad \Lambda_0 + \Sigma_{C'} \ge 0,
\ee
and,
besides,
\be
\mathrm{if}\ 
\left| - 3 \Lambda_1 + 3 \Lambda_2 - \sqrt{3}\, \Lambda_4 \right|
< 9 \Lambda_1 + 3 \Lambda_2 - 3 \sqrt{3}\, \Lambda_4,\
\mathrm{then}\ \Lambda_0 + \Sigma_{t1} \ge 0,
\ee
where
\be
\Sigma_{t1} =
\frac{4 \Lambda_1 \Lambda_2 - \Lambda_4^2}{4
  \left( 3 \Lambda_1 + \Lambda_2 - \sqrt{3}\, \Lambda_4 \right)}.
\ee

Since point $\overline{C}$ lies in $t$-path~1,
the BFB condition $\Lambda_0 + \Sigma_{\overline{C}} \geq 0$ is made redundant.

\subsubsection{$t$-path 2}

In a similar fashion to what has been done in 
the previous subsubsection,
one may connect vertices $B$,
$\overline{C'}$,
and $C$ at Eqs.~\eqref{B},
\eqref{oC'},
and~\eqref{C},
respectively,
through the \textit{Ansatz}
\be
\label{nfo2}
\chi_1 = \left( \begin{array}{c} \omega^2 \sqrt{1 + u} \\*[2mm] \sqrt{1 - u}
\end{array} \right),
\quad
\chi_2 = \left( \begin{array}{c} \omega \sqrt{1 + u} \\*[2mm] \sqrt{1 - u}
\end{array} \right),
\quad
\chi_3 = \left( \begin{array}{c} \sqrt{1 + u} \\*[2mm] \sqrt{1 - u}
\end{array} \right),
\ee
which is just \textit{Ansatz}~\eqref{nfo} with $\chi_1 \leftrightarrow \chi_2$.
Indeed,
when $u = -1$,
Eq.~\eqref{nfo2} reproduces Eq.~\eqref{B};
when $u = 0$,
it reproduces Eq.~\eqref{oC'};
and when $u = 1$,
it reproduces Eq.~\eqref{C}.

As is obvious from Eqs.~\eqref{Nxyz},
making $\chi_1 \leftrightarrow \chi_2$ is equivalent to making $z \to -z$.
In Eq.~\eqref{v4} this is equivalent to making $\Lambda_4 \to - \Lambda_4$.
So,
one must demand
\be
\Lambda_0 + \Sigma_B \ge 0, \qquad \Lambda_0 + \Sigma_C \ge 0,
\ee
and,
besides,
\be
\mathrm{if}\ 
\left| - 3 \Lambda_1 + 3 \Lambda_2 + \sqrt{3}\, \Lambda_4 \right|
< 9 \Lambda_1 + 3 \Lambda_2 + 3 \sqrt{3}\, \Lambda_4,\
\mathrm{then}\ \Lambda_0 + \Sigma_{t2} \ge 0,
\ee
where
\be
\Sigma_{t2} =
\frac{4 \Lambda_1 \Lambda_2 - \Lambda_4^2}{4
  \left( 3 \Lambda_1 + \Lambda_2 + \sqrt{3}\, \Lambda_4 \right)}.
\ee

We call $t$-path~2 the orbit-space line given
by Eqs.~\eqref{x1}--\eqref{z1} with $z \to - z$.
Since point $\overline{C'}$ lies in $t$-path~2,
the BFB condition $\Lambda_0 + \Sigma_{\overline{C'}} \geq 0$ is redundant.

\subsubsection{$t$-path 3}

One may connect vertices $C'$,
$\overline{B}$,
and $C$ through the \textit{Ansatz}
\be
\label{nfo3}
\chi_1 =
\left( \begin{array}{c} \omega^2 \sqrt{1 + u} \\*[2mm] \omega \sqrt{1 - u}
\end{array} \right),
\quad
\chi_2 =
\left( \begin{array}{c} \omega \sqrt{1 + u} \\*[2mm] \omega^2 \sqrt{1 - u}
\end{array} \right),
\quad
\chi_3 = \left( \begin{array}{c} \sqrt{1 + u} \\*[2mm] \sqrt{1 - u}
\end{array} \right).
\ee
When $u = -1$,
Eq.~\eqref{nfo3} reproduces Eq.~\eqref{C'} with the upper and lower components
of all three doublets interchanged;
when $u = 0$,
it reproduces Eq.~\eqref{oB};
and when $u = 1$,
it reproduces Eq.~\eqref{C}.
Configuration~\eqref{nfo3} produces $A_{11} = A_{22} = A_{33} = 2$
and $A_{12} = A_{23} = A_{31} = - 1 - i u \sqrt{3}$,
hence
\bs
\label{jfp3}
\ba
N &=& 6, \\
x &=& \frac{1}{3}, \label{x3} \\
x' &=& \frac{3 u^2 + 1}{12}, \label{x'3} \\
y &=& \frac{- 3 u^2 + 1}{12}, \label{y3} \\
z &=& \frac{\sqrt{3}}{6}\, u. \label{z3}
\ea
\es
We call the orbit-space line parametrically
given by Eqs.~\eqref{x3}--\eqref{z3} the $t$-path~3.

Plugging the values~\eqref{jfp3} into Eq.~\eqref{v4}
one obtains $V_4 = 12
\left( \bar a u^2 + 2 \bar k u + \bar c \right)$,
with
\bs
\ba
\bar a &=& \frac{3 \Lambda_2}{4},
\\
\bar k &=& \frac{\sqrt{3}\, \Lambda_4}{8},
\\
\bar c &=& \Lambda_0 + \frac{\Lambda_1}{4}.
\ea
\es
Accordingly,
one must demand
\be
\Lambda_0 + \Sigma_{C'} \ge 0, \qquad \Lambda_0 + \Sigma_C \ge 0,
\ee
and,
besides,
\be
\mathrm{if}\ 
\left| \Lambda_4 \right| < 2 \sqrt{3}\, \Lambda_2,\
\mathrm{then}\ \Lambda_0 + \Sigma_{t3} \ge 0,
\ee
where
\be
\Sigma_{t3} =
\frac{4 \Lambda_1 \Lambda_2 - \Lambda_4^2}{16 \Lambda_2}.
\ee

Since point $\overline{B}$ lies in $t$-path~3,
the BFB condition $\Lambda_0 + \Sigma_{\overline{B}} \geq 0$ becomes redundant.

\subsubsection{Depiction of the $t$-paths}

The $t$-paths~1,
2,
and~3 all have
\be
\label{ldp}
x = \frac{1}{3},
\qquad
x' = \frac{3 u^2 + 1}{12}
\ee
for $u \in \left[ -1, 1 \right]$,
and therefore they are all represented by the red vertical line
in Fig.~\ref{f:paths}.

\begin{figure}[h!]
  \centering
  \includegraphics[width=0.5\textwidth]{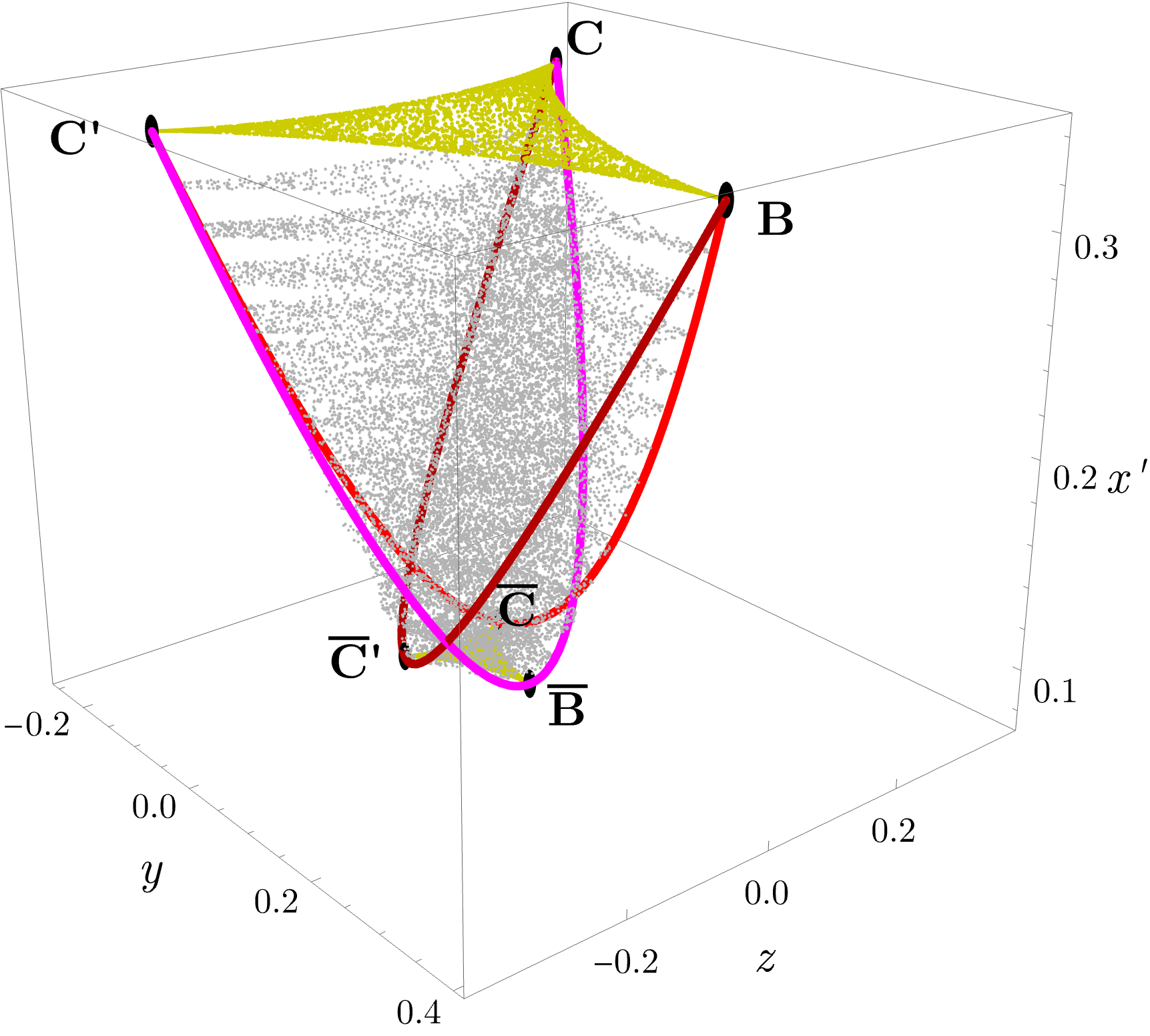}
  \caption{Orbit space projected on the $x=1/3$ hyperplane.
    Top (in yellow): the neutral configurations and the vertices $B$,
    $C$,
    and~$C'$.
    Bottom (in yellow): the vertices $\overline{B}$,
    $\overline{C}$,
    and~$\overline{C'}$.
    In grey: all the points of orbit space.
    The three $t$-paths,
    each of them joining two neutral vertices
    through a charge-breaking vertex,
    are depicted in red ($t$-path~1),
    brown ($t$-path~2),
    and pink ($t$-path~3).
}
  \label{f:slice}
\end{figure}

Figure~\ref{f:slice}
shows the orbit space projected onto the $x=1/3$ hyperplane,
\textit{cf.}\ Eq.~\eqref{ldp}.
The three $t$-paths are marked in that figure.

\subsection{The $q$-paths}

Both at point $A$ and at circumferences $\overline{D}$ and $D$
one has $A_{12} = A_{31} = 0$.
We connect those three configurations through the \textit{Ansatz}
\be
\label{ansatzOP}
\chi_1 = \left( \begin{array}{c} 0 \\ \varrho\, \sqrt{1 - u}
\end{array} \right),
\quad
\chi_2 = \left( \begin{array}{c} \sqrt{1 + u} \\ 0 \end{array} \right),
\quad
\chi_3 = \left( \begin{array}{c} e^{i \alpha} \sqrt{1 + u} \\ 0
\end{array} \right),
\ee
with $u \in \left[ -1, 1 \right]$.
In orbit space,
we call these the $q$-paths
(there is one path for each value of $\alpha$).

Configuration~\eqref{ansatzOP} has,
besides $A_{12} = A_{31} = 0$,
$A_{23} = e^{i \alpha} \left( 1 + u \right)$,
$A_{11} = 2 \left( 1 - u \right)$,
and $A_{22} = A_{33} = 1 + u$.
Hence,
\bs
\ba
N &=& 4,
\\
x &=& \frac{- 3 u^2 + 2 u + 5}{16},
\\
x' &=& \frac{u^2 + 2 u + 1}{16},
\\
y &=& x' \cos{2 \alpha},
\\
z &=& x' \sin{2 \alpha}.
\ea
\es
So,
defining
\be
\label{35}
s(\alpha) :=
\frac{\Lambda_1 \left( 1 + \cos{2 \alpha} \right)
  + \Lambda_2 \left( 1 - \cos{2 \alpha} \right)
  + \Lambda_4 \sin{2 \alpha}}{2},
\ee
it has
\be
V_4 = \frac{16}{3} \left[
  \frac{9 \Lambda_3 + 3 s(\alpha)}{16}\, u^2
  + \frac{- 6 \Lambda_3 + 6 s(\alpha)}{16}\, u
  + \Lambda_0 + \frac{\Lambda_3 + 3 s(\alpha)}{16}
  \right].
\ee
Therefore,
one must demand
\be
\label{37}
\Lambda_0 + \Sigma_A \ge 0, \qquad \Lambda_0 + \Sigma_D \ge 0,
\ee
and,
besides,
\be
\label{mcc}
\mathrm{if}\ \Lambda_3 > 0\ \mathrm{and}\ \Lambda_3 + s(\alpha) > 0,\
\mathrm{then}\ \Lambda_0 + \frac{\Lambda_3 s(\alpha)}{3 \Lambda_3
  + s(\alpha)} \ge 0.
\ee
Getting rid of the $\alpha$-dependence of conditions~\eqref{mcc}
takes some work.
Firstly, we note that,
from definition~\eqref{35},
\be
\label{nvd}
\frac{\Lambda_1 + \Lambda_2 - R}{2} \le s(\alpha)
\le \frac{\Lambda_1 + \Lambda_2 + R}{2}.
\ee
Secondly, we note that the second inequality~\eqref{mcc}
together with the \emph{negation} of the third inequality~\eqref{mcc}
are equivalent to
\be
\label{cvp}
- \Lambda_3 < s(\alpha)
< \frac{- 3 \Lambda_0 \Lambda_3}{\Lambda_0 + \Lambda_3}.
\ee
(Remember that the first condition~\eqref{37}
is precisely $\Lambda_0 + \Lambda_3 \ge 0$,
which we therefore take for granted.)
So,
when $\Lambda_3 > 0$ we need to \emph{impede} that
any $s(\alpha)$ in interval~\eqref{nvd} also is in interval~\eqref{cvp}.
One possibility of doing this is by setting
\be
\frac{- 3 \Lambda_0 \Lambda_3}{\Lambda_0 + \Lambda_3} \le - \Lambda_3
\quad \Leftrightarrow \quad
\Lambda_0 \ge \frac{\Lambda_3}{2},
\label{cvk}
\ee
which means that interval~\eqref{cvp} simply does not exist.
We concentrate on the case where inequality~\eqref{cvk} does not hold.
Then,
one must impede that the two intervals intersect.
Hence,
\be
\mathrm{if}\ \Lambda_3 > \max \left( 0, 2 \Lambda_0 \right),\ 
\mathrm{then\ either}\
\frac{\Lambda_1 + \Lambda_2 + R}{2}
\le
- \Lambda_3\
\mathrm{or}\
\frac{\Lambda_1 + \Lambda_2 - R}{2}
\ge
\frac{- 3 \Lambda_0 \Lambda_3}{\Lambda_0 + \Lambda_3}.
\ee

Since circumference $\overline{D}$ is in all
$q$-paths---it corresponds to $u=0$
in \textit{Ansatz}~\eqref{ansatzOP}---the BFB condition
$\Lambda_0 + \Sigma_{\overline{D}} \geq 0$ is
now superfluous.

\section{Conclusions and hypothesis}
\label{s:conc}

We now summarize our findings:
in order to guarantee that the SP of the $A_4$-invariant 3HDM is BFB,
one must
\begin{itemize}
\item Enforce inequalities\footnote{Conditions~\eqref{f1}--\eqref{f4}
are n\&s for the potential to be BFB
when one only uses neutral configurations
of the doublets~\cite{Ivanov:2020jra,Buskin:2021eig}.}
  \bs
  \label{f}
  \ba
  \Lambda_0 + \Lambda_3 &\ge& 0, \label{f1}
  \\
  \Lambda_0 + \Lambda_1 &\ge& 0, \label{f2}
  \\
  \Lambda_0 + \frac{\Lambda_1 + 3 \Lambda_2
    - \sqrt{3} \left| \Lambda_4 \right|}{4} &\ge& 0, \label{f3}
  \\
  \Lambda_0 + \frac{\Lambda_3}{4}
  + \frac{3 \left( \Lambda_1 + \Lambda_2 - R \right)}{8} &\ge& 0, \label{f4}
  \\
  \Lambda_0 + \frac{\Lambda_3}{4} &\ge& 0.
  \ea
  \es
\item Impede the situation where
  \bs
  \label{44}
  \ba
  \left| \Lambda_4 \right| - 2 \sqrt{3}\, \Lambda_2 &<& 0,
  \\
  \Lambda_0 + \frac{4 \Lambda_1 \Lambda_2 - \Lambda_4^2}{16 \Lambda_2} &<& 0.
  \ea
  \es
\item Forbid the situation where
  \bs
  \label{45}
  \ba
  \left| 3 \left( \Lambda_2 - \Lambda_1 \right)
  - \sqrt{3}\, \Lambda_4 \right| - 3 \left( 3 \Lambda_1 + \Lambda_2
  - \sqrt{3}\, \Lambda_4 \right) &<& 0,
  \\
  \Lambda_0 + \frac{4 \Lambda_1 \Lambda_2 - \Lambda_4^2}{4
    \left( 3 \Lambda_1 + \Lambda_2 - \sqrt{3}\, \Lambda_4 \right)} &<& 0.
  \ea
  \es
\item Forbid the situation where
  \bs
  \label{46}
  \ba
  \left| 3 \left( \Lambda_2 - \Lambda_1 \right)
  + \sqrt{3}\, \Lambda_4 \right| - 3 \left( 3 \Lambda_1 + \Lambda_2
  + \sqrt{3}\, \Lambda_4 \right) &<& 0,
  \\
  \Lambda_0 + \frac{4 \Lambda_1 \Lambda_2 - \Lambda_4^2}{4
    \left( 3 \Lambda_1 + \Lambda_2 + \sqrt{3}\, \Lambda_4 \right)} &<& 0.
  \ea
  \es
\item Discard cases where
  \bs
  \label{47}
  \ba
  \max \left( 0,\ 2 \Lambda_0 \right) - \Lambda_3 &<& 0,
  \\
  \max \left( - \Lambda_1 - \Lambda_2 - 2 \Lambda_3,\
  \Lambda_1 + \Lambda_2
  + \frac{6 \Lambda_0 \Lambda_3}{\Lambda_0 + \Lambda_3} \right)
  - \sqrt{\left( \Lambda_1 - \Lambda_2 \right)^2 + \Lambda_4^2} &<& 0.
  \hspace*{10mm}
  \ea
  \es
\end{itemize}

In our numerical work,
we have constructed one million sets of quartic couplings
$\Lambda_0$, $\Lambda_1$, $\Lambda_2$,
$\Lambda_3$, and $\Lambda_4$ that satisfy conditions~\eqref{f}.\footnote{Since
$V_4$ may be rescaled by an arbitrary positive constant,
we have uniformly sampled the quartic couplings
on a five-dimensional unit sphere.}
We checked how many of our million sets
pass the brute-force minimization procedure for $V_4$
implemented in {\tt mode 4} of the {\tt Mathematica} package 
{\tt StableWein}~\cite{Jurciukonis:2026dmc};\footnote{The mapping
between the notation used in {\tt StableWein}
and the one used here is given by
\bs\label{eq:map}
\ba
&&a_1 = a_2 = a_3 = \frac{2(\Lambda_0+\Lambda_3)}{3},\\
&&b_1 = b_2 = b_3 = \frac{2\Lambda_0-\Lambda_3}{3},\\
&&c_1 = c_2 = c_3 = \frac{\Lambda_1+\Lambda_2}{2},\\
&&d_1 = d_2 = d_3 = \left| \kappa \right|,\\
&&\epsilon = 3\arg\left( \kappa \right),
\ea\es
where
\begin{equation}
	\kappa = \tfrac12\left(\Lambda_1 - \Lambda_2 -i\Lambda_4 \right).
\end{equation}
}
we found that,
out of the one million initial sets,
957\,888 sets correspond to BFB potentials,
according to the numerical minimization procedure of {\tt StableWein}.
Now, \emph{exactly} the same 957\,888 set is obtained
when one discards,
from the original one million sets,
those where any of conditions~\eqref{44}--\eqref{47}
holds true.\footnote{The numerical work described in this paragraph
was independently performed by our colleague D.~Jur\v{c}iukonis too.
He found the same results as we did:
perfect coincidence between our necessary BFB conditions
and the numerical minimization of $V_4$.}

We therefore conjecture that \emph{our necessary conditions for BnessFB are,
in fact,
necessary and sufficient}.

\section{The case of symmetry $S_4$}
\label{s:s4}

When $\Lambda_4 = 0$ this model has an exact
$S_4$ symmetry. In that limit, our method for 
assessing BnessFB reads:
\begin{itemize}
\item Enforce inequalities
  \bs
  \label{f_S4}
  \ba
  \Lambda_0 + \Lambda_3 &\ge& 0, \label{f1_S4}
  \\
  \Lambda_0 + \Lambda_1 &\ge& 0, \label{f2_S4}
  \\
  \Lambda_0 + \frac{\Lambda_1 + 3 \Lambda_2}{4} &\ge& 0, \label{f3_S4}
  \\
  \Lambda_0 + \frac{\Lambda_3}{4}
  + \frac{3 \min \left( \Lambda_1,\ \Lambda_2 \right)}{4} &\ge& 0, \label{f4_S4}
  \\
  \Lambda_0 + \frac{\Lambda_3}{4} &\ge& 0.
  \ea
  \es
\item Impede the situation where
  \bs
  \label{44_S4}
  \ba
  \Lambda_2 &>& 0,
  \\
  \Lambda_0 + \frac{\Lambda_1}{4} &<& 0.
  \ea
  \es
\item Forbid the situation where
  \bs
  \label{45_S4}
  \ba
  \left| \Lambda_2 - \Lambda_1 \right| - 3 \Lambda_1 - \Lambda_2 &<& 0,
  \\
  \Lambda_0 + \frac{\Lambda_1 \Lambda_2}{3 \Lambda_1 + \Lambda_2} &<& 0.
  \ea
  \es
\item Discard cases where
  \bs
  \label{47_S4}
  \ba
  \max \left( 0,\ 2 \Lambda_0 \right) - \Lambda_3 &<& 0,
  \\
  \max \left( - \Lambda_1 - \Lambda_2 - 2 \Lambda_3,\
  \Lambda_1 + \Lambda_2
  + \frac{6 \Lambda_0 \Lambda_3}{\Lambda_0 + \Lambda_3} \right)
  - \left| \Lambda_1 - \Lambda_2 \right| &<& 0.
  \ea
  \es
\end{itemize}

On the other hand,
Ivanov \& Vaz\~ao~\cite{Ivanov:2020jra}
have stated that the n\&s conditions for BnessFB are
\bs
\ba
\Lambda_0 + \Lambda_1 &\ge& 0,
\\
\Lambda_0 + \Lambda_3 &\ge& 0,
\\
\Lambda_0 + \frac{\Lambda_1 + 3 \Lambda_2}{4} &\ge& 0,
\\
\Lambda_0 + \frac{\Lambda_3 + 3 \Lambda_2}{4} &\ge& 0,
\\
\Lambda_0 + \frac{\Lambda_1}{4} &\ge& 0, \label{conde}
\\
\Lambda_0 + \frac{\Lambda_3}{4} &\ge& 0, \label{condf}
\\
\Lambda_0 + \frac{\Lambda_1 + 3 \Lambda_2}{16} &\ge& 0, \label{condg}
\\
\Lambda_0 + \frac{\Lambda_3 + 3 \Lambda_2}{16} &\ge& 0; \label{condh}
\ea
\es
besides,
the situation where
\bs
\label{ev3}
\ba
\Lambda_1 - \left| \Lambda_2 \right| &>& 0,
\\
\Lambda_0 + \frac{\Lambda_1 \Lambda_2}{3 \Lambda_1 + \Lambda_2} &<& 0
\ea
\es
must be avoided,
and the situation where
\bs
\label{ev4}
\ba
\Lambda_3 - \left| \Lambda_2 \right| &>& 0,
\\
\Lambda_0 + \frac{\Lambda_2 \Lambda_3}{\Lambda_2 + 3 \Lambda_3} &<& 0
\ea
\es
must also be avoided.

We have numerically checked that the two methods
for guaranteeing BnessFB are in practice equivalent.

\vspace*{5mm}

\paragraph{Acknowledgements:} We thank Darius Jur\v{c}iukonis
for independently performing the numerical check
mentioned in section~\ref{s:conc}.
Our work was supported
by the Portuguese Foundation for Science and Technology (FCT)
through projects UIDB/00777/2020 and UIDP/00777/2020,
and by the Recovery and Resilience Plan within the scope
of investment RE-C06-i06,
measure RE-C06-i06.m02,
project 2024.01362.CERN.
The work of A.M.\ was supported by 
FCT - Fundação para a Ciência e Tecnologia, 
I.P. by project reference 2024.01340.BD and 
DOI identifier 10.54499/2024.01340.BD.

\begin{spacing}{1}
\normalsize
\bibliographystyle{JHEP}
\bibliography{a4bfb}
\end{spacing}

\end{document}